\documentclass[a4, 12in]{article}

\usepackage{blindtext}
\usepackage[a4paper, total={7in, 8in}]{geometry}

\usepackage{latexsym}
\usepackage{bm}
\usepackage{amsmath}
\usepackage{amsfonts}
\usepackage{amssymb}
\usepackage{algorithm}
\usepackage{multirow}

\usepackage{graphicx}

\usepackage{listings}
\usepackage{xcolor}

\usepackage{algorithm}
\usepackage{algpseudocode}
\usepackage{placeins}

\usepackage[hidelinks]{hyperref}
\usepackage{cleveref}

\definecolor{codegreen}{rgb}{0,0.6,0}
\definecolor{codegray}{rgb}{0.5,0.5,0.5}
\definecolor{codepurple}{rgb}{0.58,0,0.82}
\definecolor{backcolour}{rgb}{0.95,0.95,0.92}

\lstdefinestyle{mystyle}{
	backgroundcolor=\color{backcolour},   
	commentstyle=\color{codegreen},
	keywordstyle=\color{magenta},
	numberstyle=\tiny\color{codegray},
	stringstyle=\color{codepurple},
	basicstyle=\ttfamily\footnotesize,
	breakatwhitespace=false,         
	breaklines=true,                 
	captionpos=b,                    
	keepspaces=true,                 
	numbers=left,                    
	numbersep=5pt,                  
	showspaces=false,                
	showstringspaces=false,
	showtabs=false,                  
	tabsize=2
}

\usepackage{url}
\usepackage{xcolor}
\definecolor{newcolor}{rgb}{.8,.349,.1}

\usepackage{authblk}

\title{Non-Linear Super-Stencils for Turbulence Model Corrections}

\author{Jonas Luther}
\author{Patrick Jenny}
\affil{Swiss Federal Institute of Technology, Z\"urich, Switzerland}

\date{\today}

\begin{document}
	
	\maketitle
	
\begin{abstract}
Accurate simulation of turbulent flows remains a challenge due to the high computational cost of direct numerical simulations (DNS) and the limitations of traditional turbulence models. This paper explores a novel approach to augmenting standard models for Reynolds-Averaged Navier-Stokes (RANS) simulations using a Non-Linear Super-Stencil (NLSS). The proposed method introduces a fully connected neural network that learns a mapping from the local mean flow field to a corrective force term, which is added to a standard RANS solver in order to align its solution with high-fidelity data. A procedure is devised to extract training data from reference DNS and large eddy simulations (LES). To reduce the complexity of the non-linear mapping, the dimensionless local flow data is aligned with the local mean velocity, and the local support domain is scaled by the turbulent integral length scale. After being trained on a single periodic hill case, the NLSS-corrected RANS solver is shown to generalize to different periodic hill geometries and different Reynolds numbers, producing significantly more accurate solutions than the uncorrected RANS simulations.
\end{abstract}

\section{Introduction}\label{s_intro}
The numerical prediction of turbulent flows presents a challenging task. Fully resolving all scales of turbulence using direct numerical simulation (DNS) remains prohibitively expensive \cite{zienkiewicz_chapter_2014} for all but the simplest cases. Consequently, complex real-world flows, such as fully turbulent flows around airplanes, can only be treated approximately. In practice, the cheapest and most commonly used of these approximations is the Reynolds-averaged Navier-Stokes (RANS) approach~\cite{spalart_reflections_2010}. However, these methods rely on empirical turbulence models, which often require manual calibration on a case-by-case basis to produce usable results.
Data-driven techniques in turbulence modeling seek to bridge the gap between the accuracy of DNS and the efficiency of RANS simulations, shifting from traditional manual calibration, which relies on limited experimental and numerical data, to more systematic approaches that leverage the increasing availability of high-fidelity DNS data. Duraisamy et al. \cite{duraisamy_turbulence_2019} provide an extensive overview of data-driven turbulence modeling, covering traditional statistical methods, uncertainty quantification of turbulence model parameters, and the emergence of machine learning techniques. 
St\"ocker et al. \cite{stocker_dns-based_2024} present an application of the {\em SpaRTA} method to sediment-laden multi-phase flows. {\em SpaRTA} is a sparse regression technique which finds an algebraic relation between polynomial features of the mean strain and rotation rate tensors to correction terms applied to the $k$-$\varepsilon$ transport equations. This differs from the method presented in this work in three crucial ways. Most obviously, in this work only single-phase flows are considered. Second, {\em SpaRTA} only considers local flow features, while the NLSS uses non-local neighborhood information as the input for a fully connected neural network. The usage of machine learning (ML) increases the flexibility of the model at the cost of interpretability. Third, the NLSS influences the RANS solution by adding a corrective force term to the mean momentum equations, leaving the $k$-$\omega$ transport equations unchanged. 
Zhou et al. \cite{zhou_wall_2024} improve the predictive capabilities of large eddy simulations (LES) by using a fully connected neural network as a wall model. Specifically, they sample different mean flow features at three equidistant points along the normal of each wall cell to predict the wall shear stress and the single parameter of a modified mixing length model used to determine the eddy viscosity in these wall cells. In contrast to our sampling procedure, their points are sampled at a fixed distance unaffected by turbulent scales. 
Quattromini et al. \cite{quattromini_operator_2023} and Zhou et al. \cite{zhou_frame-independent_2022} present alternative methods of incorporating non-local information into their predictions. The former use a graph neural network (GNN) \cite{scarselli_graph_2009} to predict the Reynolds stress tensor in the entire domain, while the latter use a vector-cloud neural network, which instead of sampling mean fields using interpolation, directly takes in the positions and data of neighboring cells as an input. In contrast to our work, both of these approaches fully bypass the classical turbulence model.
Ling et al. \cite{Ling_Kurzawski_Templeton_2016} correct eddy-viscosity models by predicting anisotropic corrections of the Reynolds stress tensor using a Galilean-invariant tensor basis network, whose input features are also non-dimensionalized using values predicted by the base turbulence model. However, they only use local information for these predictions.
Boureima et al. \cite{BOUREIMA2022110924} take a slightly different but related approach: Instead of modeling a correction force, they calibrate parameters of the non-local BHR Reynolds stress model by reformulating the RANS solver itself as a neural network - allowing existing automatic differentiation frameworks to optimize these parameters using backpropagation. 
Overall, the literature shows a clear trend towards leveraging machine learning to improve turbulence modeling, with various approaches focusing on local corrections (St\"ocker et al.~\cite{stocker_dns-based_2024} and Ling et al.~\cite{Ling_Kurzawski_Templeton_2016}), non-local information (Quattromini et al.~\cite{quattromini_operator_2023} and X. Zhou et al.~\cite{zhou_frame-independent_2022}), wall modeling (Zhou et al.~\cite{zhou_wall_2024}), and parameter calibration (Boureima et al.~\cite{BOUREIMA2022110924}). Certainly, the combination of machine learning with traditional turbulence models holds significant promise for improving predictive capabilities in computational fluid dynamics (CFD), especially in cases where RANS models struggle to capture complex flow physics accurately.
\\ \ \\
In this paper, data-driven corrections are applied to empirical turbulence models to obtain more accurate predictions without requiring manual intervention. Specifically, the Non-Linear Super-Stencil (NLSS) is introduced. 
That it is impossible to unambiguously describe the state of turbulence at some point only based on local mean quantities is well known. Traditionally, non-localness is introduced through additional partial differential equations accounting, e.g. for transport of turbulent kinetic energy and turbulent frequency, or through elliptic relaxation models. In contrast,
the NLSS is designed to learn a non-linear mapping from the neighboring mean flow, sampled on a large stencil, to a corrective force term in the mean momentum equation. To minimize the required training data for learning this mapping, the input and output spaces are reduced in a physically informed manner: First, the super-stencil is rotated and scaled to align with the local mean velocity and turbulent length scale. Second, the sampled mean flow values are transformed to dimensionless features using their respective characteristic turbulent scales derived from the turbulence model.
\\ \ \\
The premises for the present work are that in principle it is possible to (i) correct a cheap low fidelity RANS model by probing the neighboring mean flow, (ii) that only relatively small neighborhood domains have to be considered, and (iii) that the non-linear relation between model correction and neighboring mean flow data can be learned from available high fidelity data via machine learning. The NLSS based model correction presented in this paper builds on these premises by introducing a method that uses non-local, physically informed sampling and dimensionless feature transformations to improve the generalizability and accuracy of RANS simulations. Note that the correction does not need to "see" the whole flow, but only a small local window. Thus, as long as the local patterns are "known", the trained NLSS correction in principle should  generalize to cases with different global flow topologies. So far, as proof of concept, we have shown that this is the case for different periodic hill geometries with different Reynolds numbers, but more work is required to investigate and generalize the approach for three dimensional mean flows, compressible flows, transition, etc.
\\ \ \\
The remainder of this paper is structured as follows: Next, the closure problem is stated, then the Non-Linear Super-Stencil is introduced, which includes its training and application for RANS simulations, and in section~\ref{s_numerical_experiments} numerical experiments are presented. Finally, the paper closes with a discussion (section~\ref{s_concl}) and a methods section (section~\ref{a_methods}), providing detailed information regarding implementation, neural network type and architecture. 
\\ \ \\
Incompressible flow with constant density $\rho$ and kinematic viscosity $\nu$ is considered. The goal is to compute the mean velocity $\bar{\bm u}$ and the mean pressure $\bar{p}$ at low computational cost by solving the Reynolds averaged continuity and momentum equations
\begin{eqnarray}
\label{e_cont}
\frac{\partial \bar u_i}{\partial x_i}&=&0
\textrm{\ \ \ and}\\
\label{e_rans}
\frac{\partial \bar u_i}{\partial t}+\frac{\partial \bar u_i\bar u_j}{\partial x_j}
&=&
-\frac{1}{\rho}\frac{\partial \bar p}{\partial x_i}
+
\frac{\partial}{\partial x_j}\left(2\nu\bar{S}_{ij}\right)
-
\frac{\partial\overline{u'_i u'_j}}{\partial x_j},
\end{eqnarray}
respectively, where $\bar{S}_{ij}=0.5(
{\partial\bar u_i/\partial x_j}+{\partial\bar u_j}/{\partial x_i}
)$ is the mean rate of strain.
Note that the Reynolds stresses $\overline{u'_i u'_j}$ in the Reynolds averaged Navier Stokes (RANS) equation~\eqref{e_rans} require modeling. Here closure is achieved via the Boussinesq eddy viscosity approximation~\cite{Boussinesq-1877}
\begin{eqnarray}\label{e_edv}
-\overline{u'_i u'_j}
&\approx&
2\nu_t\bar{S}_{ij}
\ -\ 
\frac{2}{3}k\delta_{ij}
,
\end{eqnarray}
where $k=\overline{u'_i u'_i}/2$ is the turbulent kinetic energy and $\nu_t$ the eddy viscosity. In the current implementation, the $k$-$\omega$ model \cite{wilcox_turbulence_1998} is considered and the eddy viscosity is approximated as $\nu_t=k/\omega$. This implies that the two additional model equations 
\begin{eqnarray}
\label{e_k}
\frac{\partial k}{\partial t}+\frac{\partial \bar u_jk}{\partial x_j}&=&
\frac{\partial}{\partial x_i}\left(
(\nu+\alpha_k\nu_t)\frac{\partial k}{\partial x_i}
\right)
\ +\ 
2\nu_t\bar{S}_{ij}\bar{S}_{ij}
\ -\ 
\beta^* k\omega
\textrm{\ \ \ and\ \ \ }\\
\label{e_omega}
\frac{\partial \omega}{\partial t}+\frac{\partial \bar u_j\omega}{\partial x_j}&=&
\frac{\partial}{\partial x_i}\left(
(\nu+\alpha_\omega\nu_t)\frac{\partial \omega}{\partial x_i}
\right)
\ +\ 
2\nu_t\bar{S}_{ij}\bar{S}_{ij}
\ \frac{\gamma\omega}{k}
\ -\ 
\beta\omega^2
\end{eqnarray}
for $k$ and the turbulent frequency $\omega$ have to be solved. The standard model parameter values are taken from from \cite{wilcox_turbulence_1998} and can be found in \cref{t_parameter_values}.
Note that other eddy viscosity models could be employed instead; the $k$-$\omega$ model was chosen here due to its simplicity and its relatively good performance near walls.
By substituting the eddy viscosity assumption \eqref{e_edv} into Eq.~\eqref{e_rans} one obtains the modeled RANS equation
\begin{eqnarray}
\label{e_rans_modeled}
\frac{\partial \bar u_i}{\partial t}+\frac{\partial \bar u_i\bar u_j}{\partial x_j}
&=&
-\frac{1}{\rho}\frac{\partial p_e}{\partial x_i}
+
\frac{\partial}{\partial x_j}\left(2\nu_e\bar{S}_{ij}\right)
+
f_i,
\end{eqnarray}
where $p_e=\bar p+2\rho k/3$ is the effective pressure, $\nu_e=\nu+\nu_t$ the effective viscosity and 
\begin{eqnarray}
\label{e_f}
f_i
&=&
-\frac{\partial\overline{u'_i u'_j}}{\partial x_j}
\ +\ 
\frac{2}{3}\frac{\partial k}{\partial x_i}
\ -\ 
\frac{\partial}{\partial x_j}\left(2\nu_t\bar{S}_{ij}\right)
\end{eqnarray}
a correction force accounting for the model error.
Note that if one knew the correction force $f_i$, then the exact mean velocity field $\bar{\bm u}$ could be computed by solving the closed system composed of Eqs.~\eqref{e_cont}, \eqref{e_k}, \eqref{e_omega} and \eqref{e_rans_modeled} with an appropriate numerical scheme (for this work a cell centered finite volume method was employed; see section \ref{aa_rans_solver}). The worthwhile question thus is, whether there exists a general approach to obtain an accurate estimate of $f_i$ based on the computed mean fields. Note that the identical question can be asked for any other Reynolds stress closure.
\\ \ \\
Envisioned is a general approach to accurately estimate the RANS model correction force $\bm f$ in Eq.~\eqref{e_rans_modeled} based on the computed mean fields. Therefore we hypothesize that there exists an unambiguous, non-linear mapping that allows to express the correction force at any point in the domain as a function of the mean flow pattern in a neighborhood of that point. And further, that this map can be learned from empirical training data  using a neural network.
\\ \ \\
In order to quantify the flow pattern around a point $\bm x^*$, a compact super stencil aligned with the mean velocity $\bar{\bm u}^*$ is considered (the superscript $*$ denotes quantities evaluated at point $\bm x^*$). The stencil points can be denoted as
\begin{eqnarray}
\label{e_stencilpoints}
\bm x^*_{I,J,K}&=&
c_ls^*_l\left(
\frac{I}{n_1}\bm e^*_1+\frac{J}{n_2}\bm e^*_2+\frac{K}{n_3}\bm e^*_3
\right)
\textrm{\ \ \ for }
I,J,K\in\{-n_1,...,n_1\}\times\{-n_2,...,n_2\}\times\{-n_3,...,n_3\}
,
\end{eqnarray}
where $s_l^*=\sqrt{k^*}/\omega^*$ is the integral length scale, $\bm e^*_1=\bar{\bm u}^*/|\bar{\bm u}^*|$ the dimensionless unit vector in flow direction, $\bm e^*_3$ a unit vector perpendicular to $\bm e^*_1$ with 
$\min_{\bm e^*_3}(|(\bm\nabla|\bar{\bm u}|)^*\cdot\bm e^*_3|)$, and $\bm e^*_2$ a unit vector perpendicular to both $\bm e^*_1$ and $\bm e^*_3$. The parameter $c_l$ controls the size of the stencil support (all algorithmic parameter values used for the numerical experiments in this paper are given in \cref{t_parameter_values}), while $n_{1}$ , $n_{2}$ and $n_{3}$ define the number of stencil points; note that in two dimensions $n_3=0$. 
\begin{figure}[htbp] 
	\centering
	\begin{tabular}{c}
		\includegraphics[width=3in]{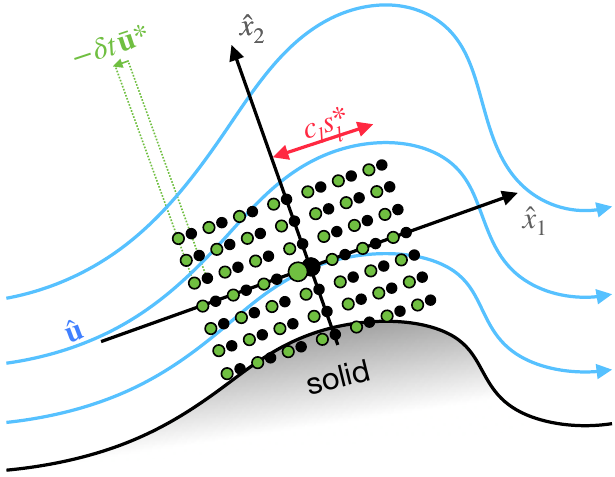}
	\end{tabular}
	\caption{
		Schematic of a two dimensional super-stencil (black dots) with $n_1=n_2=3$. The stencil is centered around the point $\bm x^*=\bm x^*_{0,0,0}$, and it is aligned with the mean velocity $\bar{\bm u}^*$ at its center.
		For completeness' sake, the shifted stencil used for Galilean transformation (green dots shifted by $-\delta t\bar{\bm u}^*$ with respect to the black dots) is also shown. This is explained in \cref{a_neural_network}.
	}
	\label{f_stencil}
\end{figure}
The black dots in \cref{f_stencil} depict a two dimensional super-stencil with $n_1=n_2=3$. At each stencil point the mean velocity, its rate of change, the mean strain rate, the ratio of eddy viscosity to effective viscosity and a solid indicator $s\in\{\textrm{True, False}\}$ (indicating, whether the stencil point is beyond a solid boundary or not) are sampled. In order to reduce the variability, all quantities are normalized by the reference length $s_l^*$, the time $1/\omega^*$ and the velocity $\sqrt{k^*}$. Further, rotation and Galilean transformation are applied to obtain a transformed coordinate system which is aligned with $\bar{\bm u}^*$ and which moves with $\bar{\bm u}^*$. The normalized transformed quantities then are
\begin{eqnarray}
\hat{\bm x}&=&\bm R^T({\bm x}-{\bm x}^*)\frac{\omega^*}{\sqrt{k^*}},\\
\label{e_stencil_u}
\hat{\bm u}&=&\bm R^T(\bar{\bm u}-\bar{\bm u}^*)\frac{1}{\sqrt{k^*}},\\
\label{e_stencil_dotu}
\hat{\dot{\bm u}}
&=&
\bm R^T\left(\frac{\partial\bar{\bm u}}{\partial t}+(\bar{\bm u}^*\cdot\nabla)\bar{\bm u}\right)\frac{1}{\omega^*\sqrt{k^*}},\\
\hat{\bm S}&=&\bm R^T\bar{\bm S}\bm R\frac{1}{\omega^*}
\end{eqnarray}
and
\begin{eqnarray}
\label{e_stencil_f}
\hat{\bm f}^*&=&\bm R^T\bm f^*\frac{1}{\omega^*\sqrt{k^*}},
\end{eqnarray}
where $\bm R$ is a rotation tensor to align the local coordinate system with $\bar{\bm u}^ *$.
Note that normalization, aligning the local frame of reference with the mean velocity, and exploiting Galilean invariance is crucial to reduce the variability of flow patterns to be learned. 
The designated task now is to express $\hat{\bm f}^*$ at point $\bm x^*$ as function of $\hat{\bm u}(\hat{\bm x}_{I,J,K})$, $\hat{\dot{\bm u}}(\hat{\bm x}_{I,J,K})$, $\hat{\bm S}(\hat{\bm x}_{I,J,K})$, $q(\hat{\bm x}_{I,J,K})=\nu_t(\hat{\bm x}_{I,J,K})/\nu_e(\hat{\bm x}_{I,J,K})$ and $s(\hat{\bm x}_{I,J,K})$ (note that $\hat{\bm x}_{I,J,K}={\bm x}^*_{I,J,K}/s_l^*$ with $I,J,K\in\{-n_1,...,n_1\}\times\{-n_2,...,n_2\}\times\{-n_3,...,n_3\}$ are the normalized and transformed stencil point coordinates). 
A more detailed description of the implementation to obtain these transformed normalized stencil point values is provided in section \ref{a_neural_network}. 
Instead of trying to express such a mapping via traditional mathematical techniques, e.g. via algebraic expressions and additional coupled partial differential equations, machine learning using a neural network is considered. For training high fidelity mean flow fields are required.
\\ \ \\
To obtain training data, Eqs.~\eqref{e_cont}, \eqref{e_k} and \eqref{e_omega} together with 
\begin{eqnarray}
\label{e_rans_modeled_relax}
\frac{\partial \bar u_i}{\partial t}+\frac{\partial \bar u_i\bar u_j}{\partial x_j}
&=&
-\frac{1}{\rho}\frac{\partial p_e}{\partial x_i}
+
\frac{\partial}{\partial x_j}\left(2\nu_e\bar{S}_{ij}\right)
+
\underbrace{\chi(\bar{u}_i^\textrm{high fidelity}-\bar{u}_i)}_{\approx f_i},
\end{eqnarray}
are numerically solved; details about the implementation are given in sections~\ref{aa_rans_solver} and algorithm~\ref{alg:training_data}. There are multiple reasons why the force term $f_i$ is introduced as a relaxation source term:

\begin{itemize}
	\item The direct evaluation of $\bm f$ from training data, as given by Eq.~\eqref{e_rans_modeled}, involves solving Eqs.~\eqref{e_k} and \eqref{e_omega} given a mean reference high-fidelity velocity field. This was found to be numerically undesirable. Even when a consistent solution was reached, the presence of interpolation artifacts in the reference velocity field led to noisy force fields.

	\item In contrast, the approximation $f_i\approx\chi(\bar{u}_i^\textrm{high fidelity}-\bar{u}_i)$ allows the solution of Eq. \eqref{e_rans_modeled_relax} to deviate slightly from the high fidelity solution, for instance, to alleviate continuity errors. This leads to a smoother reference solution of $k^\text{ref}$,  $\omega^\text{ref}$, and $\bar{\bm{u}}^\text{ref}$, and by extension, a smoother correction force $\bm{f}^\text{ref}$. The tradeoff between exact reconstruction and regularity is controlled by the relaxation rate $\chi$ and further discussed below.
	
	\item The approach is inspired by recent variational data assimilation (DA) approaches (see Brenner et al. \cite{BRENNER2024117026} and Plogmann et al. \cite{PLOGMANN2024117052}) in the context of RANS and URANS, which also rely on reference mean velocity data (in their case spatially sparse) to reconstruct an entire RANS solution. In their methods, the tradeoff between regularization and faithful reconstruction is even more pronounced.
	
	\item Finally, this approach allows to vary the relaxation rate $\chi$ across the domain, which here is mainly used to blend out the correction force very close to walls (see also \cref{f_q}), where the turbulent intensity is low.
\end{itemize}
Note that while $\chi$ is an algorithmic parameter, $f_i=\lim_{\chi\rightarrow\infty}\chi(\bar{u}_i^\textrm{high fidelity}-\bar{u}_i)$, that is, one has to ensure that $\chi$ is large enough, but at the same time not too large in order to have a smoothing effect (more details can be found in section \ref{aa_rans_solver}). Here it was set to $\chi=\chi_\textrm{max}\min(2q,1)$, where $q=\nu_t/\nu_e$ is the ratio of eddy viscosity to effective viscosity (its behavior is discussed in section~\ref{a_neural_network}; see also \cref{f_q}). All algorithmic parameter values used for the numerical experiments in this paper are given at the beginning of section~\ref{sec:params}. Using this expression for $\chi$  has the additional desired effect of blending out the correction force in regions of low turbulent intensity, e.g. near walls. Once the fields $\bar{\bm u}$, $q$ and $\bm f$ are computed, one can easily obtain sets of corresponding $\hat{\bm f}^*$, $\hat{\bm u}(\hat{\bm x}_{I,J,K})$, $\hat{\dot{\bm u}}(\hat{\bm x}_{I,J,K})$, $\hat{\bm S}(\hat{\bm x}_{I,J,K})$, $q(\hat{\bm x}_{I,J,K})$ and $s(\hat{\bm x}_{I,J,K})$ values at many points $\bm x^*$ within the domain and use them to train the NLSS, i.e., a neural network (see \ref{a_neural_network}). For example, in two dimensions with $n_1=n_2=7$, the neural network needs to have 2025 input- and 2 output nodes.
\\ \ \\
The final objective, of course, is to employ the NLSS during RANS simulations in order to obtain the model correction force $\bm f$. Therefore, for each grid node $\bm x^*$ the stencil points $\bm x^*_{I,J,K}$ have to be determined according to Eq.~\eqref{e_stencilpoints}. The input to the trained neural network then consists of all values $\hat{\bm u}$, $\hat{\dot{\bm u}}$, $\hat{\bm S}$, $q$ and $s$ at these stencil points, and the output is $\hat{\bm f}^*$. The transformation $\hat{\bm f}^*\rightarrow {\bm f}^*$ to obtain the correction force at $\bm x^ *$ is straight forward. More details are provided in section~\ref{a_neural_network} and algorithm~\ref{alg:inference}.

\section{Results}\label{s_numerical_experiments}
For all simulations presented in this paper, the algorithmic parameter values specified in table~\ref{t_parameter_values} were used. Note, however, that so far no systematic investigation of our method's sensitivity regarding these algorithmic parameters has been conducted, that is, the proposed values are based on heuristic experimentation.
%

\subsection{Test Case Geometries, Boundary- and Flow Conditions}
Periodic hills are standard benchmark cases in computational fluid dynamics with a wide range of available experimental and numerical high fidelity data from both DNS and LES. In this paper, mean flows obtained from two sets of numerical simulations are used as reference data. The first set is based on a range of DNS performed by Xiao et al. \cite{xiao_flows_2020} for different hill geometries at a fixed Reynolds number of $\mathrm{Re}=h_\textrm{crest}u_\textrm{bulk}/\nu=5600$ ($h_\textrm{crest}$ denotes the crest hight and $u_\textrm{bulk}$ the bulk velocity at the crest location). The second set consists of data obtained by Gloerfelt et al. \cite{gloerfelt_large_2019} from three LES with varying Reynolds numbers of $\mathrm{Re}\in\{2800, 10595, 19000\}$ at a fixed hill geometry.
\begin{figure}[htbp] 
	\centering
	\begin{tabular}{c}
		\includegraphics[width=4in]{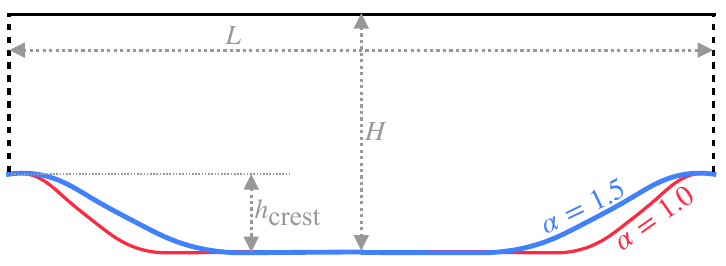}
	\end{tabular}
	\caption{
		Schematic depicting the geometries of the periodic hill cases. The parameters shown are the domain length $L$, the domain height $H$, the crest height $h_\textrm{crest}$ and the hill stretch factor $\alpha$. The dashed lines represent the periodic in- and outflow boundaries, and solid lines represent walls.
	}
	\label{f_geometry}
\end{figure}
The different hill geometries are characterized by the domain height $H$, the crest hight $h_\textrm{crest}$, the hill stretch factor $\alpha$ and the domain length $L$; see \cref{f_geometry}. The former two parameters are fixed to $H=3.036$m and $h_\textrm{crest}=1$m. The hill profile depends on $\alpha$ and is defined as a piecewise cubic spline, whose definition is provided as {\em Python}-code in section~\ref{aa_rans_solver}. 
The in- and outflow boundary conditions at the crest of the hill are periodic, while the top and bottom boundaries represent walls. Walls are represented by no-slip Dirichlet boundary conditions for the velocity, and default conditions for the turbulent quantities $k$ and $\omega$ as described in \cite{liu2016thorough}.
The pressure is constrained by Neumann boundary conditions at the walls,  while it is fixed to a specified value at a single reference point. To ensure a consistent flow rate through the domain, the bulk velocity at the hill crest is constrained to $u_\textrm{bulk}=1$m/s.

\subsection{Training}

\begin{table}[h!]
	\centering
	\begin{tabular}{lcc||r|rrr}
		\hline
		\textbf{case name} & \textbf{usage} & \textbf{reference} & \textbf{grid cells} & $\alpha$ & $L$  & $\mathrm{Re}$ \\
		\hline
		\hline
		\texttt{case 1}& testing & DNS & 18000 & 1.0 & 6m & 5600 \\
		\texttt{case 2}& testing & DNS & 36000 & 1.0 & 12m & 5600 \\
		\texttt{case 3}& testing & DNS & 32700 & 1.5 & 10.929m & 5600 \\
		\texttt{case 4}& testing & DNS & 41700 & 1.5 & 13.929m & 5600 \\
		\hline
		\texttt{case 5}& testing & DNS & 27000 & 1.0 & 9m & 5600 \\
		\texttt{case 6}& \textbf{training} & LES & 27000 & 1.0 & 9m & 10595 \\
		\texttt{case 7}& testing & LES & 27000 & 1.0 & 9m & 19000 \\
		\hline
	\end{tabular}
	\caption{Cases used for training and testing with geometric parameters, Reynolds number and number of grid cells used by the finite volume method for the RANS simulations. Channel- and crest heights are kept fix at $H=3.036$m and $h_\textrm{crest}=1$m, respectively. Cases~ 1-4 \cite{xiao_flows_2020} all share the same Reynolds number, while cases~ 5-7 \cite{gloerfelt_large_2019} all share the same geometry.}
	\label{t_cases}
\end{table} The NLSS was trained and validated on a set of the available reference data summarized in table~\ref{t_cases}.
For training only a single case - namely case 6 - was employed. This is to demonstrate that with very little training data it is possible to capture the non-linear dependencies which are required to correct the RANS model for flows with different geometries and Reynolds numbers. The reference forces were obtained as detailed in section~\ref{s_intro}. For each cell of the reference solution, a super-stencil and its mirrored twin were sampled to create a training dataset of $54000$ NLSS maps. The neural network was trained over $200$ epochs, with the hyper-parameters chosen as described in section~\ref{a_neural_network}. On an AMD RX 7900 XTX, a high-end consumer GPU, training took about a minute. To monitor the model's accuracy during this process, the model was validated on all cases (1-7) after each epoch.

\subsection{Cross Case Validation}
\label{sec:results}
\begin{figure}[htbp]
	\centering
	\includegraphics[width=6in]{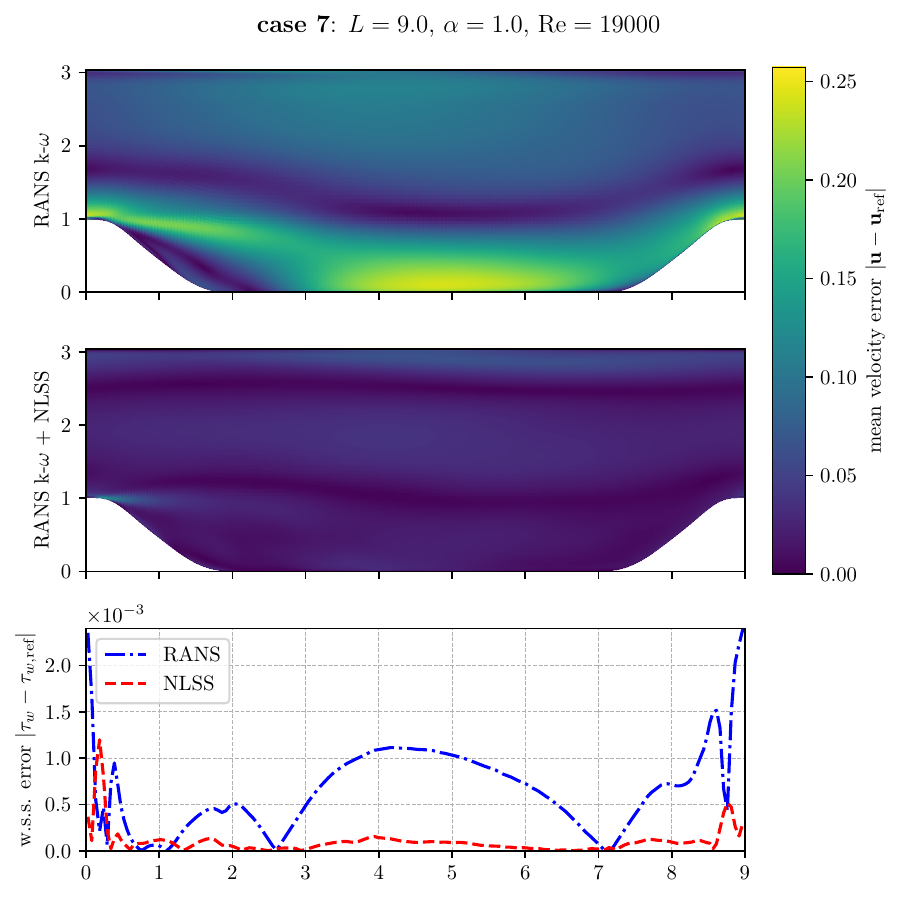}
	\caption{Error analysis of test case 7: Velocity error maps of the uncorrected (top row) and NLSS-corrected (middle row) RANS $k$-$\omega$ model. The bottom row shows the absolute wall shear stress errors (w.r.t. the reference data) along the bottom wall of the uncorrected (blue) and NLSS-corrected (red) solutions.}
	\label{fig:error_case7}
 \end{figure}

\begin{figure}[htbp]
	\centering
	\includegraphics[width=5.5in]{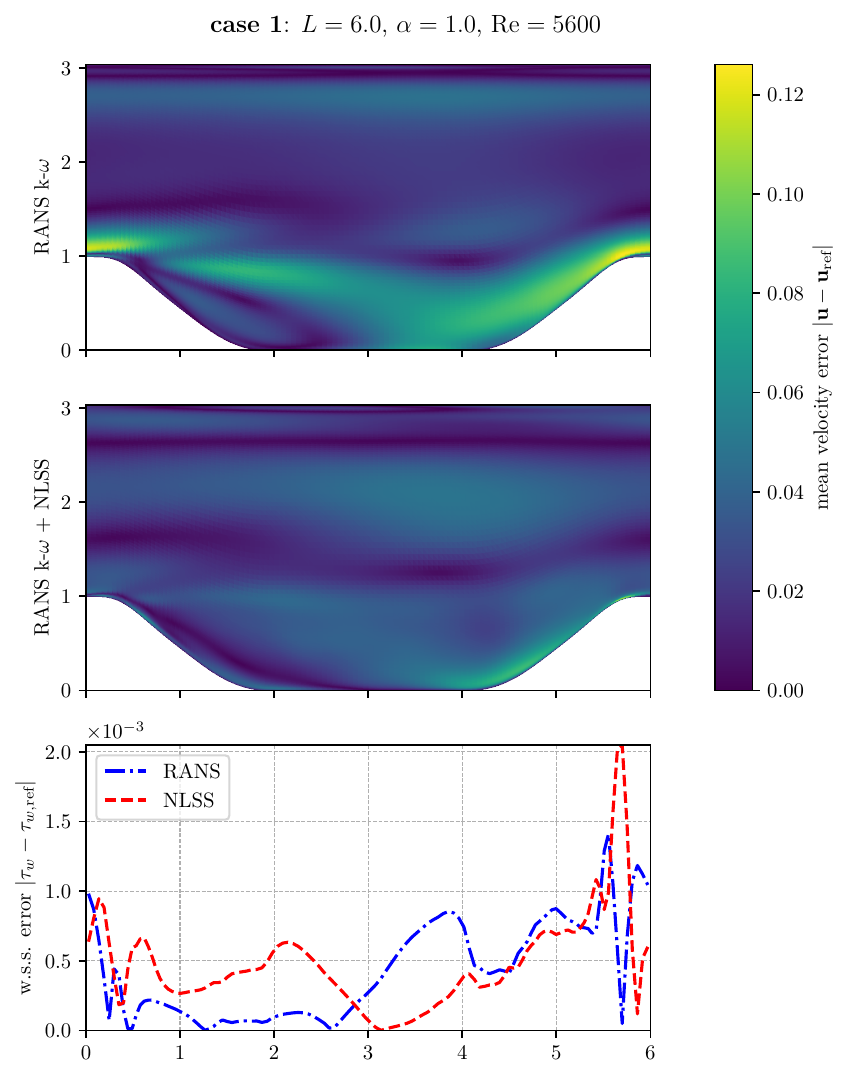}
	\caption{Error analysis of test case 1 (worst accuracy achieved in our study): Velocity error maps of the uncorrected (top row) and NLSS-corrected (middle row) RANS $k$-$\omega$ model. The bottom row shows the absolute wall shear stress errors (w.r.t. the reference data) along the bottom wall of the uncorrected (blue) and NLSS-corrected (red) solutions.}
	\label{fig:error_case1}
 \end{figure}

\begin{figure}[htbp] 
	\centering
	\begin{tabular}{c}
		\includegraphics[width=6.5in]{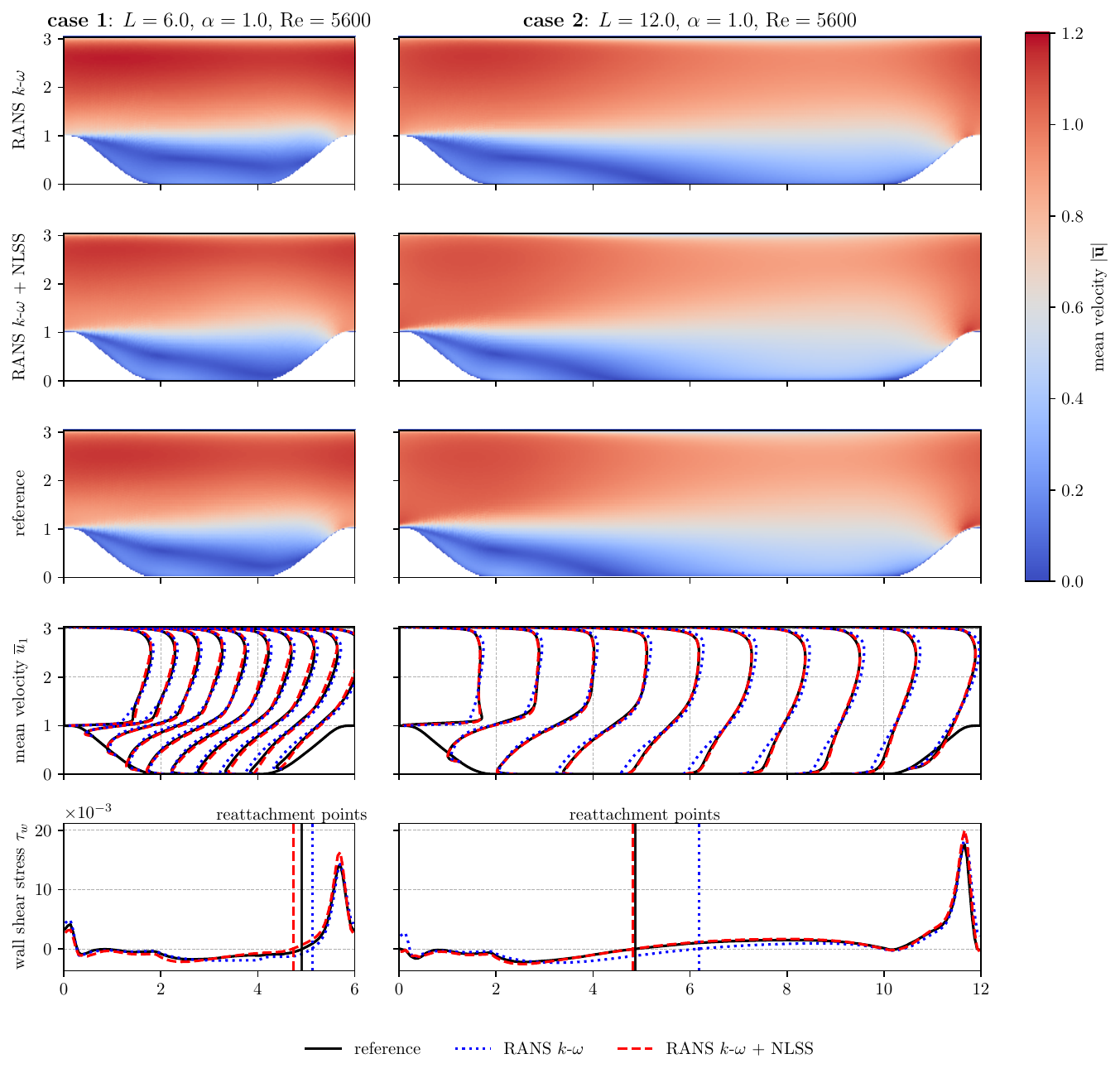}
	\end{tabular}
	\caption{
		RANS simulation results of test cases~1 (left) and 2 (right) with and without NLSS-correction along with high fidelity data. Mean velocity magnitude maps: Uncorrected RANS model (first row),  NLSS-corrected RANS model (second row), reference (third row). Corresponding profiles of the horizontal mean velocity component at ten different downstream locations are shown in the fourth row, and the wall shear stresses along the bottom wall of the periodic hill geometries are shown in the bottom row. Locations of mean flow re-attachment of reference and NLSS-corrected solutions are marked by the solid black and dashed red vertical lines; those  predicted by the uncorrected RANS simulations by the blue dotted vertical lines.
	}
	\label{f_result_u_1_2}
\end{figure}

\begin{figure}[htbp] 
	\centering
	\begin{tabular}{c}
		\includegraphics[width=6.5in]{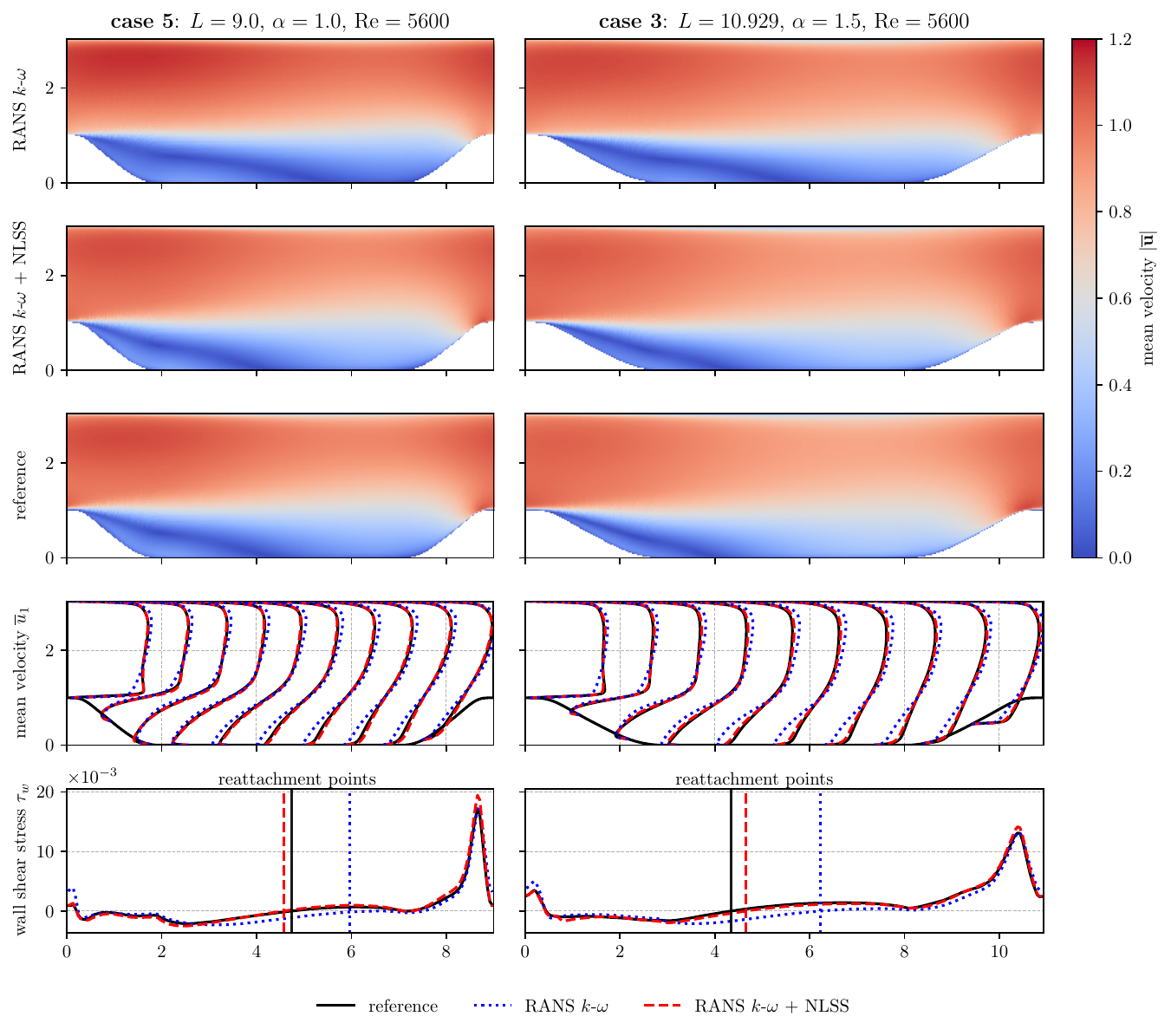}
	\end{tabular}
	\caption{
		RANS simulation results of test cases~5 (left) and 3 (right) with and without NLSS-correction along with high fidelity data. Mean velocity magnitude maps: Uncorrected RANS model (first row),  NLSS-corrected RANS model (second row), reference (third row). Corresponding profiles of the horizontal mean velocity component at ten different downstream locations are shown in the fourth row, and the wall shear stresses along the bottom wall of the periodic hill geometries are shown in the bottom row. Locations of mean flow re-attachment of reference and NLSS-corrected solutions are marked by the solid black and dashed red vertical lines; those  predicted by the uncorrected RANS simulations by the blue dotted vertical lines.
	}
	\label{f_result_u_3_5}
\end{figure}

\begin{figure}[htbp] 
	\centering
	\begin{tabular}{c}
		\includegraphics[width=3.5in]{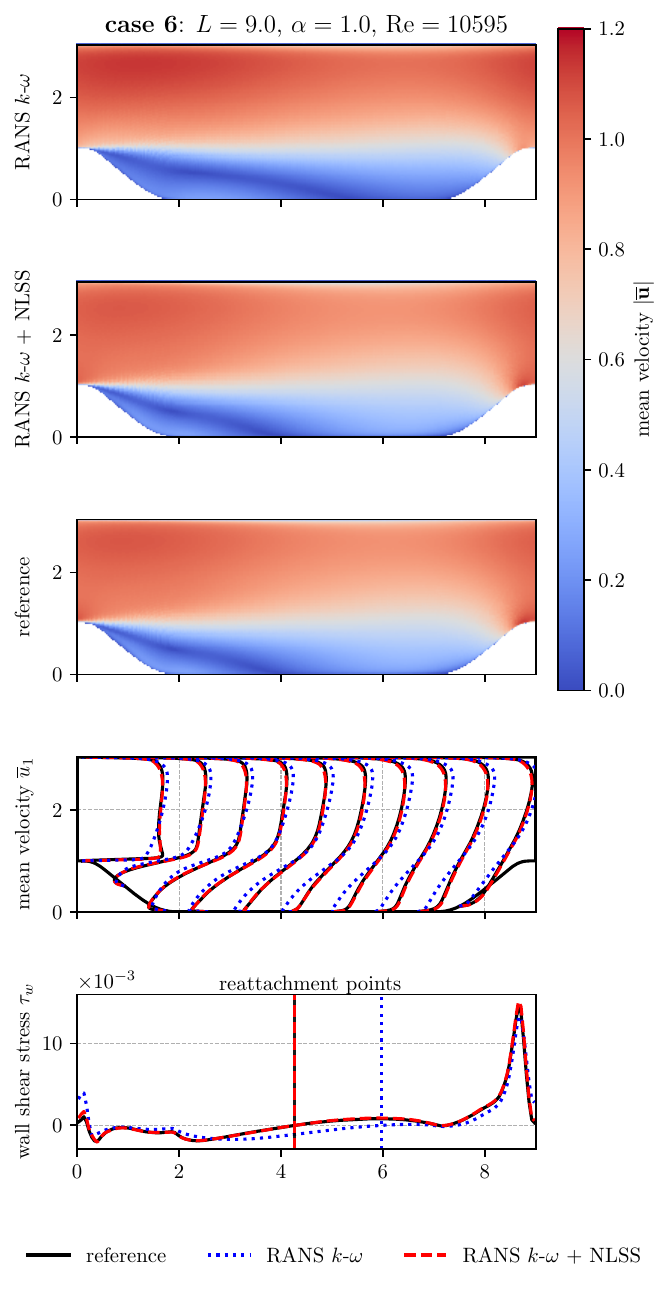}
	\end{tabular}
	\caption{
		RANS simulation results of training case~6 with and without NLSS-correction along with high fidelity data. Mean velocity magnitude maps: Uncorrected RANS model (first row),  NLSS-corrected RANS model (second row), reference (third row). Corresponding profiles of the horizontal mean velocity component at ten different downstream locations are shown in the fourth row, and the wall shear stresses along the bottom wall of the periodic hill geometries are shown in the bottom row. Locations of mean flow re-attachment of reference and NLSS-corrected solutions are marked by the solid black and dashed red vertical lines; that  predicted by the uncorrected RANS simulation by the blue dotted vertical line.
	}
	\label{f_result_u_6}
\end{figure}

The trained model was validated for all cases listed in table~\ref{t_cases}. First, uncorrected RANS simulations were conducted with the {\em OpenFOAM} \cite{Weller1998} solver \texttt{simpleFOAM} and the $k$-$\omega$ model \cite{wilcox_turbulence_1998}. The obtained solutions then were used as initial conditions for the NLSS-corrected RANS simulations. More details on the implementation and numerical parameters can be found in sections~\ref{aa_rans_solver} and \ref{sec:params}. 
\begin{figure}[htbp] 
	\centering
	\begin{tabular}{c}
		\includegraphics[width=6.5in]{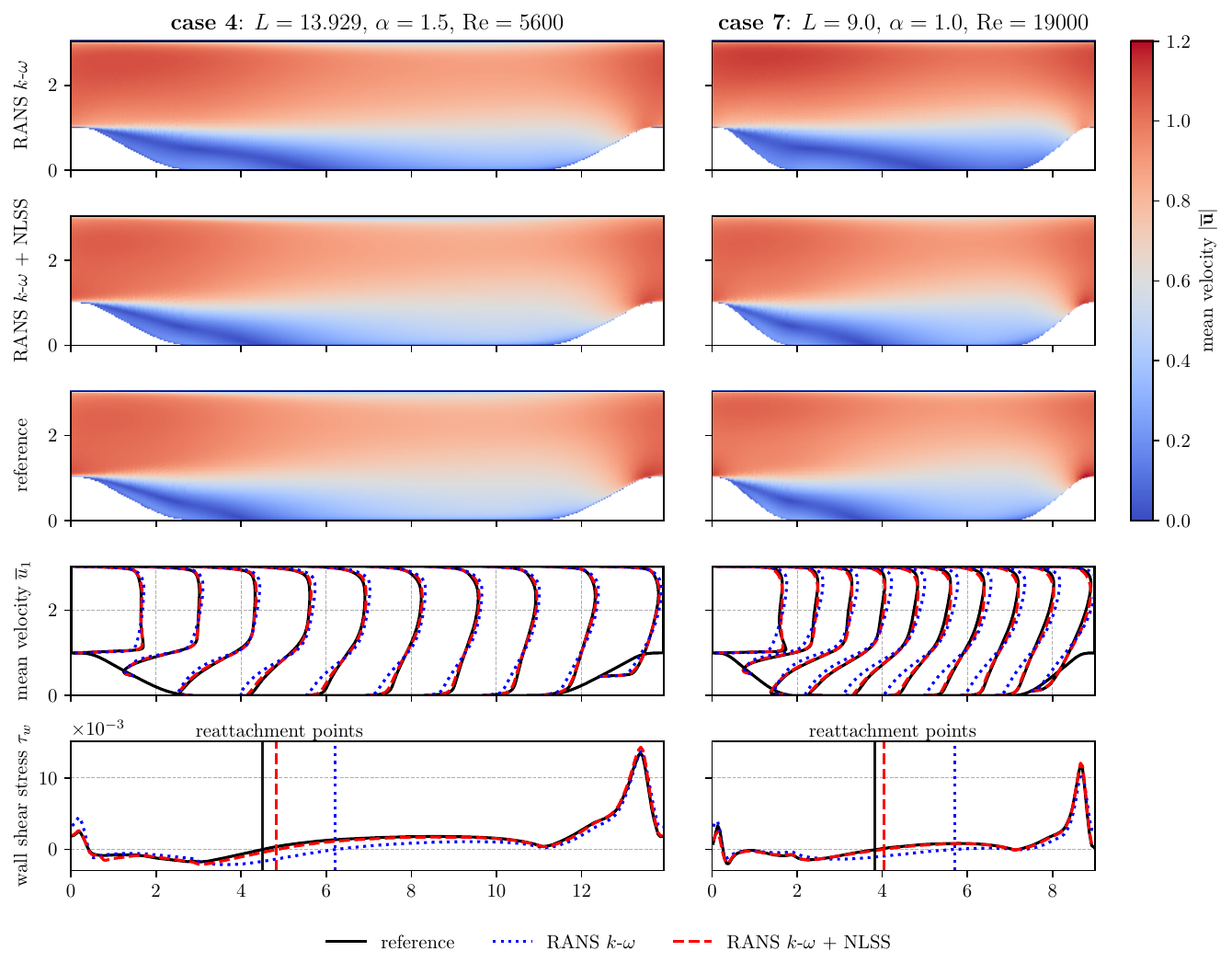}
	\end{tabular}
	\caption{
		RANS simulation results of test cases~4 (left) and 7 (right) with and without NLSS-correction along with high fidelity data. Mean velocity magnitude maps: Uncorrected RANS model (first row),  NLSS-corrected RANS model (second row), reference (third row). Corresponding profiles of the horizontal mean velocity component at ten different downstream locations are shown in the fourth row, and the wall shear stresses along the bottom wall of the periodic hill geometries are shown in the bottom row. Locations of mean flow re-attachment of reference and NLSS-corrected solutions are marked by the solid black and dashed red vertical lines; those  predicted by the uncorrected RANS simulations by the blue dotted vertical lines.
	}
	\label{f_result_u}
\end{figure}
\begin{figure}[htbp] 
	\centering
	\begin{tabular}{c}
		\includegraphics[width=6.5in]{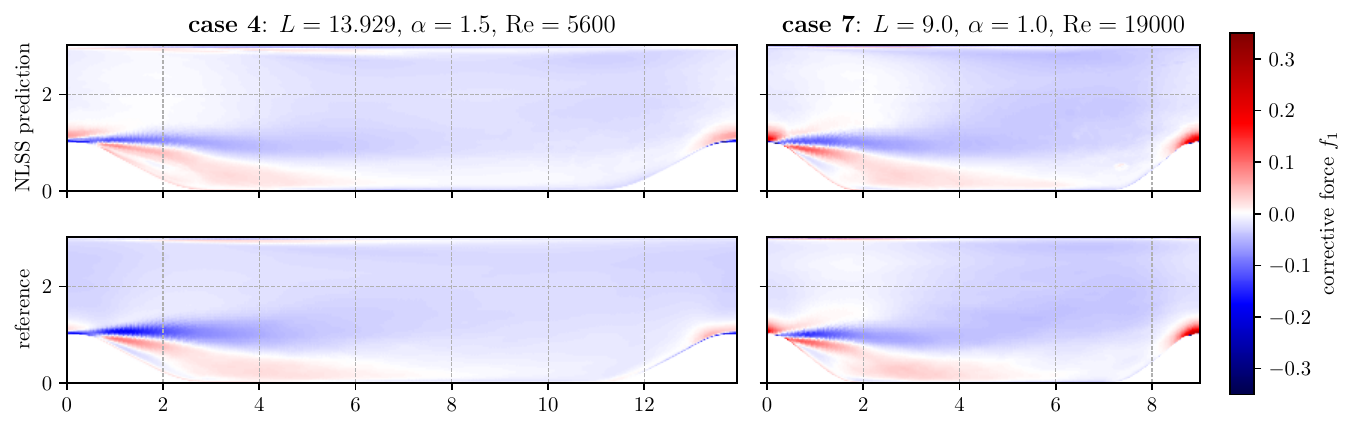}
	\end{tabular}
	\caption{
		Horizontal component $f_1$ of the model correction force for the test cases~4 (left) and 7 (right); in the top row as predicted by the NLSS and in the bottom row as extracted from the high fidelity reference data (see section~\ref{s_intro}).
	}
	\label{f_result_f}
\end{figure}
\\ \ \\
As shown in table~\ref{t_cases}, the cases are grouped into two sets; the first one (cases~1-4) considering different geometries and the second one (cases~5-7) different Reynolds numbers. The respective high fidelity references for the following comparisons are taken from Xiao et al.~\cite{xiao_flows_2020} and Gloerfelt et al. \cite{gloerfelt_large_2019}.
Figure~\ref{f_result_u} shows RANS simulation results of test cases~4 (left) and 7 (right) with and without NLSS correction along with high fidelity data. Note that the NLSS was trained only with the data from case~6 for a domain length of $L=9$m, a hill stretch factor of $\alpha=1$ and a Reynolds number of $Re=10595$, while $L=13.929$m, $\alpha=1.5$ and $Re=5600$ were chosen for case~4 and $L=9$m, $\alpha=1$ and $Re=19000$ for case~7. The top three rows in \cref{f_result_u} depict mean velocity magnitude maps, that is, the uncorrected RANS model results in the first, the NLSS-corrected RANS model results in the second, and the reference fields in the third row. Corresponding profiles of the horizontal mean velocity component at ten different downstream locations are shown in the fourth row  (solid, red dashed and blue dotted lines represent reference data, NLSS-corrected and uncorrected RANS model results, respectively), and the bottom row shows the wall shear stresses along the bottom wall of the periodic hill geometries. For both test cases it can be observed that the NLSS-corrected RANS solutions closely agree with the reference data (much better than the uncorrected ones), thus demonstrating the ability of the NLSS to generalize across different geometries and Reynolds numbers. How dramatic the improvements of the RANS results due to the NLSS-correction are is clearly illustrated by the locations of mean flow re-attachment (where the wall shear stress switches from negative to positive values), which are marked by the vertical lines in \cref{f_result_u} (solid black lines for the reference, dashed red lines for the NLSS-corrected, and dashed blue lines for the uncorrected RANS solutions). In both test cases the locations predicted by the NLSS-corrected RANS simulations are in very close agreement with the reference, while the re-attachment points predicted by the uncorrected RANS simulations (vertical dashed blue lines) are found much further downstream.
\\ \ \\
Figure~\ref{f_result_f} shows the horizontal component of the model correction force for the test cases ~4 (left) and 7 (right); in the top row as predicted by the NLSS and in the bottom row as extracted from the high fidelity reference data (see section~\ref{s_intro}). The model correction force is most pronounced in regions with high mean strain rate and/or large pressure gradients, i.e., above the crest of the hill and around the edge of the separation bubble. Despite both cases being unseen, the NLSS model manages to reproduce the reference force remarkably well. The slight mismatch above the separation bubble on the left side of test case~4 does not seem to have a large effect on the resulting mean velocity. In contrast, the correction force for test case~7 is reproduced almost perfectly (also see \cref{fig:error_case7}). 
\begin{table}[h]
	\centering
	\label{t_errors}
	\caption{Relative L2-norms of the mean velocity ($\bar{\bm u}$) prediction errors versus reference data. The numbers show a general trend of the NLSS model significantly improving the RANS solution.}
	\vspace{0.5em}
	\begin{tabular}{l|r|r}
	\hline
	\textbf{case} & \multicolumn{2}{c}{\textbf{error}} \\
	\hline
	& $\bar{\bm u}_\text{RANS}$ & $\bar{\bm u}_\text{NLSS}$\\
	\hline
	\hline
	\texttt {case 1} & 0.152 & 0.123 \\
	\texttt {case 2} & 0.332 & 0.083 \\
	\texttt {case 3} & 0.461 & 0.097 \\
	\texttt {case 4} & 0.457 & 0.105 \\
	\hline
	\texttt {case 5} & 0.281 & 0.090 \\
	\texttt {case 6} & 0.396 & 0.067 \\
	\texttt {case 7} & 0.515 & 0.128 \\
	\hline
	\end{tabular}
\end{table}
\\ \ \\
\Cref{t_errors} shows the L2-norms of the mean velocity prediction errors of the uncorrected and NLSS-corrected RANS $k$-$\omega$ models, integrated over the domain. As also qualitatively shown in \cref{f_result_u,f_result_u_1_2,f_result_u_3_5,f_result_u_6}, these numbers again indicate that the NLSS correction significantly improves the accuracy of the $k$-$\omega$ model. As expected, the model performs best on the case it was trained on (training case 6). Note however, that even in this case some generalization is involved: Despite being only trained on the reference solution, the NLSS correction is able to predict forces which steer the simulation in the correct direction; starting from the initial \emph{uncorrected} RANS solution, which is not represented in the training data. 
\\ \ \\
We attribute the deviations between NLSS-corrected RANS solution and reference data observed in test case~1 to the fact that in contrast to the other cases the reattachment occurs at a strongly curved surface, which is a local flow pattern not present in the employed training data. This interpretation is supported by \cref{fig:error_case1}, which indicates that the main error in mean velocity prediction is found at the upstream side of the hill, while the error at the crest is smaller in comparison. We expect that training the NLSS correction on a richer set of local flow patterns will also improve the accuracy for test case~1, but this is subject of future research. 
\begin{table}[h]
	\centering
	\caption{Runtime comparisons between the uncorrected and NLSS corrected RANS simulations for each test case.}
	\vspace{1.0em}
	\begin{tabular}{l|c|c}
	\hline
	\textbf{case} & \multicolumn{2}{c}{\textbf{runtime}}\\
	\hline
	 & $t_\text{RANS}$ & $t_\text{NLSS}$ \\
	\hline
	\hline
	\texttt{case 1} & 1 min 30 sec & 54 min 7 sec \\
	\texttt{case 2} & 3 min 44 sec & 98 min 33 sec \\
	\texttt{case 3} & 4 min 16 sec & 86 min 29 sec \\
	\texttt{case 4} & 5 min 38 sec & 109 min 7 sec \\
	\hline
	\texttt{case 5} & 3 min 14 sec & 77 min 34 sec \\
	\texttt{case 6} & 3 min \ 8 sec & 75 min 37 sec \\
	\texttt{case 7} & 3 min 33 sec & 76 min 40 sec \\
	\hline
	\end{tabular}
	
	\label{t_runtime}
\end{table}
\\ \ \\
 As can be seen in \cref{t_runtime}, the computational efficiency of our current implementation still has room for improvement; in all cases, the NLSS-corrected simulations took considerably longer than their RANS counterparts. Most of this overhead is due to the interpolation needed to extract the stencil point values, and we are confident that this can be optimized and the runtime significantly reduced.


\section{Discussion}\label{s_concl}
This paper presents a novel approach to augmenting traditional RANS-based turbulence models with machine learning, specifically through the Non-Linear Super-Stencil (NLSS). The NLSS was implemented within the framework of {\em OpenFOAM}, employing a fully connected neural network to capture non-linear dependencies of RANS model corrections based on mean flow data. The model's input and output features are aligned with the local flow velocity and made dimensionless using the local values of $k$ and $\omega$ provided by the turbulence model. This procedure greatly reduces the amount of required training data compared to learning the same non-linearity in a fully-dimensional space. The model’s ability to generalize from a single periodic hill training case to a range of other test cases supports this hypothesis.
\\ \ \\
While the method presented here generalizes promisingly well within the family of periodic hills, it has yet to be tested on other benchmark cases such as flow around airfoils or channel flows with different geometries. Future work should involve testing on a broader range of cases to fully assess the NLSS's robustness, adaptability and generality. 
\\ \ \\
It is also worth noting that the implementation presented here serves as a proof-of-concept. There exist a myriad of possible optimizations, such as different stencil geometries, the usage of more advanced neural network architectures such as graph neural networks (GNNs), the usage of more involved turbulence models, a more rigorous selection of the hyper-parameters mentioned in this paper, and of course more training. 
\\ \ \\
Besides investigating the NLSS-based model correction across different families of turbulent flow cases, its use for many more non-linear physical phenomena could be explored. Examples include rarefied gas flows, turbulent combustion, and even large-scale weather and climate simulations, where different NLSS could provide corrections using  non-dimensionalized, non-local features. In summary, the basic NLSS-concept is very general and can be applied to many more problems involving closure models and partial differential equations.

\FloatBarrier
\section{Methods}\label{a_methods}
Next, details of the RANS solver and required changes thereof, geometries and computational grids, interpolation of high fidelity data onto the RANS grids, integration of the NLSS into the RANS solver for training and simulation runs, and of the neural network used in combination with the NLSS are provided. 

\subsection{RANS Solver, Geometries, Gridding and Interpolation}
\label{aa_rans_solver}
To compute steady state solutions of the coupled Eqs.~\eqref{e_cont}, \eqref{e_k}, \eqref{e_omega} and \eqref{e_rans_modeled}, the standard {\em Semi-Implicit Method for Pressure-Linked Equations} (SIMPLE) solver \cite{patankar1983calculation} \texttt{simpleFOAM} provided by {\em OpenFOAM} \cite{Weller1998} is employed. As described in section~\ref{s_intro}, the Boussinesq eddy viscosity approximation with the $k$-$\omega$ model by Wilcox \cite{wilcox_turbulence_1998} for closure of the eddy viscosity $\nu_t$ is used. Note that in Eqs.~\eqref{e_k} and \eqref{e_omega} standard model constants are used (see \cref{t_parameter_values}).
\\ \ \\
{\em Correction force:}\\
To get the model correction force $\bm f^*$ for a grid cell with center $\bm x^*$, periodically (e.g. every tenth iteration) the NLSS points $\bm x^*_{I,J,K}$ are determined as described by Eq.~\eqref{e_stencilpoints}. Then the transformed normalized quantities $\hat{\bm u}$, $\hat{\dot{\bm u}}$, $\hat{\bm S}$, $q$ and $s$ are evaluated at all these stencil points (see section~\ref{s_intro}) and fed into the trained neural network, which outputs the transformed normalized model correction force $\hat{\bm f}^*$. In principle, the non-normalized force $\bm f^*=\omega^*\sqrt{k^*}\bm R\hat{\bm f}^*$ then can be used to correct the divergence of the modeled Reynolds stress tensor at the stencil center $\bm x^*$, but it is favorable to employ its divergence free part $\bm f^\textrm{df}=\bm f^*-\bm\nabla\phi$ (Helmholtz projection) instead, where the potential $\phi$ is computed by solving the Poisson equation
\begin{eqnarray}
\frac{\partial^2\phi}{\partial x_i\partial x_i}&=&\frac{\partial f^*_i}{\partial x_i}.
\end{eqnarray}
Also for training the divergence free field $\bm f-\bm\nabla\phi$ instead of $\bm f=\chi(\bar{\bm u}^\textrm{high fidelity}-\bar{\bm u})$ (see Eq.~\eqref{e_rans_modeled_relax}, which is considered instead of Eq.~\eqref{e_rans_modeled} for training) is employed.
\\ \ \\
{\em Damping for stabilization:}\\
Although the solver \texttt{simpleFOAM} is designed for computing steady state solutions, in some occasions low frequency oscillations in the mean flow solutions remained. Therefore, the damping term 
\begin{eqnarray}
\label{e_damp}
\chi^\textrm{damp}(\bar{\bm u}^\textrm{MA}-\bar{\bm u})
\end{eqnarray}
was added to the right-hand-side of Eq.~\eqref{e_rans_modeled} during prediction simulations (note that this has nothing to do with the relaxation term in Eq.~\eqref{e_rans_modeled_relax}), where
\begin{eqnarray}
\bar{\bm u}^{\textrm{MA},n+1}
&=&
\mu^\textrm{mem}\bar{\bm u}^{\textrm{MA},n}
\ +\ 
(1-\mu^\textrm{mem})\bar{\bm u}^{n+1}
\end{eqnarray}
is the moving average of $\bar{\bm u}$
with the memory factor $\mu^\textrm{mem}$ and the damping rate $\chi^\textrm{damp}$; the superscripts $n$ and $n+1$ denote old and new iteration levels, respectively. Note that damping has no effect on the solution, if steady state is reached (since in that case $\bar{\bm u}^\textrm{MA}=\bar{\bm u}$). During the transient phase, however, the proposed damping term proved to damp the undesired low frequency oscillations.
Optimal values of $\mu^\textrm{mem}$ and $\chi^\textrm{damp}$ must be sufficiently high to suppress limit cycles, but too large values would slow down the solver by inhibiting its ability to diverge from the initial conditions (all algorithmic parameter values used for the numerical experiments in this paper are given in table~\ref{t_parameter_values}). 
\\ \ \\
{\em Geometries and computational grids:}\\
For creating the hill geometry dependent on the domain length $L$ and the shape factor $\alpha$, splines are used; the implementation is found in the  following Python script:
\begin{lstlisting}[language=Python, caption={\em Python}-code to create the bottom wall geometry of the periodic hill cases using the length and shape parameters as input \cite{xiao_flows_2020}.]
import numpy as np

def hill_spline(x):
  x *= 28
  h = 0.0
  if 0 <= x < 9:
    h = np.minimum(28., 2.8e+01 + 0.0e+00 * x + 6.775070969851e-03
    *x**2-2.124527775800e-03*x**3) 
  elif x < 14:
    h = 2.507355893131E+01 + 9.754803562315E-01 * x -
    1.016116352781E-01*x**2+1.889794677828E-03*x**3
  elif x < 20:
    h =   2.579601052357E+01 + 8.206693007457E-01 * x -
    9.055370274339E-02*x**2+1.626510569859E-03*x**3 
  elif x < 30:
    h = 4.046435022819E+01 - 1.379581654948E+00 * x +
    1.945884504128E-02*x**2-2.070318932190E-04*x**3
  elif x < 40:
    h =   1.792461334664E+01 + 8.743920332081E-01 * x -
    5.567361123058E-02*x**2+6.277731764683E-04*x**3 
  elif x <= 54:
    h = np.maximum(0., 5.639011190988E+01 - 2.010520359035E+00 * x 	
    +1.644919857549E-02*x**2+2.674976141766E-05*x**3)
  return h / 28.0

def hill_profile(x, alpha, L):
  HILL_WIDTH = 3.85714285714 * alpha
  if x < 0.5 * HILL_WIDTH:
    return hill_spline(x / alpha)
  elif x < length - 0.5 * HILL_WIDTH:
    return 0
  else:
    return hill_spline((L - x) / alpha)

# example usage
L=9 
alpha = 1.0
x = np.linspace(0, L, 100)
y = hill_profile(x, alpha, L)
\end{lstlisting}
All RANS simulations presented in this paper were performed on structured wall-fitted grids generated by the \texttt{blockMesh} utility provided by {\em OpenFOAM}. All grids comprise $150$ cells in  vertical direction and $20 L/H$ cells in  horizontal direction. To keep the vertical extent of the first cells near walls in the vicinity of $y^+ \approx 1$, the grids are refined toward the top and bottom walls with a grading factor of $20$.
\\ \ \\
{\em Interpolation of data between grids:}\\
The interpolation procedure from the DNS/LES grids to the RANS grids differs for the two sets of simulations (cases~1-4 and cases 5-7, respectively). The DNS were performed on Cartesian, non-wall conforming meshes using an immersed boundary method to account for the hill profile. The provided mean flow data was obtained by averaging over time and in transverse direction. Since the RANS grids are wall conforming and have a different resolution, artifacts in the near-wall regions can arise when the mean DNS fields are interpolated onto the RANS grids. With the LES grids, which are wall conforming and employ the same near wall resolution in normal direction as the RANS grids, interpolation artifacts are much smaller. Concretely, to interpolate reference data from fine DNS or LES grids to the RANS grids, \texttt{LinearNDInterpolator} from the \texttt{scipy} \cite{2020SciPy-NMeth} {\em Python}-package is fit to the fine grid and then evaluated at the RANS grid points.

\subsection{Non-Linear Super-Stencil (NLSS) and Neural Network}
\label{a_neural_network}
To integrate the {\em Python}-based neural network into the {\em C++}-based {\em OpenFOAM}-solver, \texttt{PyTorch} is employed, which provides two key functionalities, i.e., (i) the module \texttt{torch.jit.script} \cite{devito2022torchscript}, which enables cross-language serialization of trained neural networks, and (ii) \texttt{libtorch}, the {\em C++} front-end for \texttt{PyTorch}. Concretely, the neural network is trained in {\em Python}, serialized to a checkpoint file, and loaded by the solver using the \texttt{libtorch} front-end, thus allowing it to predict forces on the fly during simulations. 
\\ \ \\
{\em Sampling:}\\
The first step in sampling the mean flow with the non-linear super-stencil is to determine the mesh cells of each stencil point. For this task, {\em OpenFOAM} provides the \texttt{meshSearch} class. To accelerate the sampling process, the cell indexes returned by the \texttt{meshSearch} class are precomputed at the beginning of the simulation and cached in a fine regular grid that covers the entire domain. Using \texttt{OpenMP}, this grid is accessed in parallel, allowing rapid sampling of mean flow values at each stencil point for each cell in the domain. Since the stencil points do not necessarily coincide with the cell centers, the mean flow values are linearly interpolated using \texttt{interpolateCellPoint}.
\\ \ \\
{\em Data normalization, transformation and augmentation:}\\
After sampling, the mean flow values are transformed and normalized to obtain the dimensionless input features $\hat{\bm u}$, $\hat{\dot{\bm u}}$, $\hat{\bm S}$, $q$ and $s$; see section~\ref{s_intro}.
The sampled velocity is transformed to the stencil's reference frame by first subtracting the reference velocity $\bar{\bm u}^*$ (Galilean transformation). The result is then multiplied from the left by ${\bm R}^T$, where the orthogonal matrix ${\bm R}$ defines the rotation from stencil coordinates to global coordinates. As a Galilean-invariant rank-2 tensor, the mean strain rate $\bar{\bm S}$ transforms as ${\bm R}^T\bar{\bm S}{\bm R}$. Both quantities are then non-dimensionalized by their respective characteristic scales $\sqrt{k^*}$ and $\omega^*$. The indicators $q$ and $s$ are not transformed, since they  already are dimensionless and rotation- as well as Galilean invariant. Transformation and normalization are crucial to ensure that the input space to the neural network is as small as possible, which in turn reduces the number of required training samples. In other words, scale-invariance of the Navier-Stokes equations is exploited in order to simplify the learning task. 
Due to Galilean transformation the rate of change of the velocity has to be transformed as well; see Eq.~\eqref{e_stencil_dotu}. That is, instead of $\partial\bar{\bm u}/\partial t$ in the global frame of reference, $\partial\bar{\bm u}/\partial t+(\bar{\bm u}^*\cdot\bm\nabla)\bar{\bm u}$ has to be considered. The current implementation is based on the approximation
\begin{eqnarray}
\left(\frac{\partial\bar{\bm u}}{\partial t}
\ +\ 
(\bar{\bm u}^*\cdot\bm\nabla)\bar{\bm u}\right)_{{\bm x},t}
&\approx&
\frac{\bar{\bm u}({\bm x},t)-\bar{\bm u}({\bm x-\bar{\bm u}^*\delta t},t-\delta t)}{\delta t},
\end{eqnarray}
and by substituting above expression into Eq.~\eqref{e_stencil_dotu} one obtains
\begin{eqnarray}
\hat{\dot{\bm u}}({\bm x},t)
&\approx&
\frac{\hat{\bm u}({\bm x},t)}{\delta t}
\ -\ 
\bm R^T\left(
\bar{\bm u}({\bm x-\bar{\bm u}^*\delta t},t-\delta t)
\right)
\frac{1}{\delta t\omega^*\sqrt{k^*}}.
\end{eqnarray}
This clearly shows that $\bm R^T\left(
\bar{\bm u}({\bm x-\bar{\bm u}^*\delta t},t-\delta t)
\right)
(\delta t\omega^*\sqrt{k^*})^{-1}$ provides the same additional information as $\hat{\dot{\bm u}}({\bm x},t)$, if $\delta t$ is very small. Here we chose $\delta t=c^\textrm{lag}/\omega^*$ (all algorithmic parameter values used for the numerical experiments in this paper are given in table~\ref{t_parameter_values}). Thus, and since it is straight forward to extract 
$\bar{\bm u}^\textrm{lag}({\bm x},t)=\bar{\bm u}({\bm x-\bar{\bm u}^*\delta t},t-\delta t)$
from the computed mean velocity field, $\hat{\bm u}^\textrm{lag}=\bm R^T
\bar{\bm u}^{\textrm{lag}}
(\omega^*\sqrt{k^*})^{-1}$ is used here instead of $\hat{\dot{\bm u}}$ for training of the neural network. Concretely, an additional stencil shifted by the distance $-c^\textrm{lag}\bar{\bm u}^*/\omega^*$ (as depicted by the green dots in \cref{f_stencil}) with respect to the original one (black dots in \cref{f_stencil}) is introduced, and at all shifted stencil points the mean velocities are evaluated at time $t-\delta t$ and then transformed analogously as the mean velocities at the original stencil points; see Eq.~\eqref{e_stencil_u}. 
Note that for steady state computations the values can be evaluated at $t$, but in the general case the shifted stencil point values have to be interpolated in time using also the mean velocity field from the previous time step. In summary, in the current implementation, instead of $\{\hat{\bm u}, \hat{\dot{\bm u}}, \hat{\bm S}, q, s\}$, the set $\{\hat{\bm u}, \hat{\bm u}^\textrm{lag}, \hat{\bm S}, q, s\}$ of normalized and transformed stencil point values provides the input to the neural network. Note that $q=\nu_t/(\nu_t+\nu)\in[0,1]$ provides indirect information about the Reynolds number to the neural network. For example, near walls it approaches zero, and in highly turbulent regions it asymptotically reaches one. 
\begin{figure}[htbp] 
	\centering
	\begin{tabular}{c}
		\includegraphics[width=5.5in]{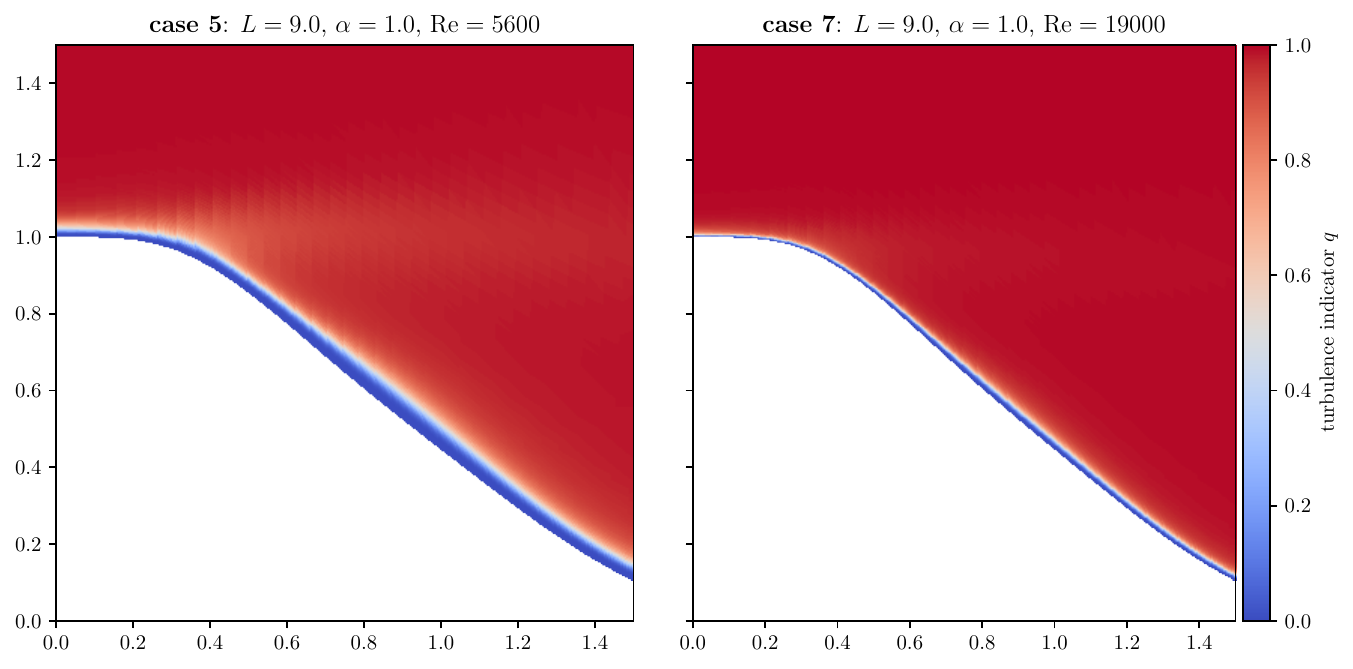}
		\	\end{tabular}
	\caption{
		Maps of $q$ in a sub-region of cases~5 (left) and 7 (right).
	}
	\label{f_q}
\end{figure}
For illustration, the left and right maps in \cref{f_q} depict $q$ in a sub-region of cases~5 and 7, respectively. Note the region of very low $q$-values is much smaller for case~7 (right plot) with a Reynolds number of $Re=19000$ compared to case~5 (left plot) with a Reynolds number of $Re=5600$.
\\ \ \\
In addition to normalization and transformation, the training data is augmented by also considering the mirrored twins. Concretely, for every training stencil with the stencil values $\hat{\bm u}, \hat{\bm u}^\textrm{lag}, \hat{\bm S}, q$ and $s$, a mirrored stencil with the stencil point values 
\begin{eqnarray}
\hat{\bm u}^\textrm{mirrored}(\hat{\bm x}_{I,J,K})&=&{\bm H}\hat{\bm u}(\hat{\bm x}_{I,-J,K}), \\
\hat{\bm u}^\textrm{lag,mirrored}(\hat{\bm x}_{I,J,K})&=&{\bm H}\hat{\bm u}^\textrm{lag}(\hat{\bm x}_{I,-J,K}), \\
\hat{\bm S}^\textrm{mirrored}(\hat{\bm x}_{I,J,K})&=&{\bm H}\hat{\bm S}(\hat{\bm x}_{I,-J,K}){\bm H},\\
q^\textrm{mirrored}(\hat{\bm x}_{I,J,K})&=&q(\hat{\bm x}_{I,-J,K}),\\
s^\textrm{mirrored}(\hat{\bm x}_{I,J,K})&=&s(\hat{\bm x}_{I,-J,K})
\textrm{\ \ \ and\ \ \ }\\
\hat{\bm f}^\textrm{*,mirrored}&=&{\bm H}\hat{\bm f}^*
\end{eqnarray}
is added to the training data, where
\begin{eqnarray}
\bm H&=&
\left[
\begin{array}{ccc}
1&0&0\\
0&-1&0\\
0&0&1
\end{array}
\right]
\end{eqnarray}
is the mirroring operator. Note that this only accounts for mirroring at the $\hat{x}_1$-$\hat{x}_3$-plane, which suffices for two dimensional cases. For three dimensional cases, in addition mirroring at the $\hat{x}_1$-$\hat{x}_2$-plane can be considered for further training data augmentation.
\\ \ \\
{\em Neural network type and architecture:}\\
To model the non-linear dependence of the force term on the input features, a deep neural network is used. 
\begin{figure}[htbp] 
	\centering
	\begin{tabular}{c}
		\includegraphics[width=2in]{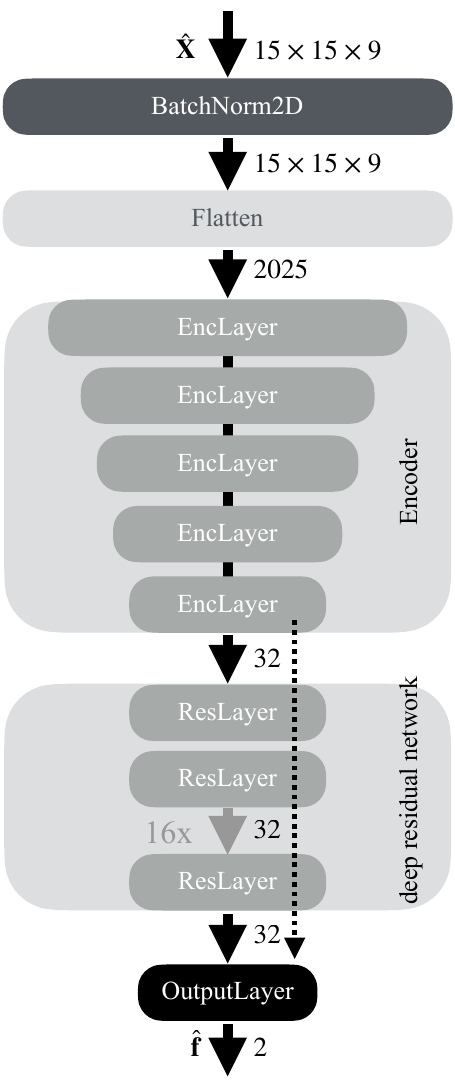}
	\end{tabular}
	\caption{
		Schematic of the fully connected neural network architecture used in the numerical experiments. The network consists of an encoder with five layers, followed by a series of 16 residual layers and a final output layer. The dashed connection indicates the residual connection, allowing direct information flow from the encoder to the output layer.
	}
	\label{f_neural_network}
\end{figure}
As shown in \cref{f_neural_network}, the network consists of two parts, i.e., of (i) an encoder, which reduces the dimensionality of the input features, followed by (ii) a deep residual network able to capture more complex non-linear dependencies \cite{hornik_multilayer_1989}. The encoder consists of 5 fully connected layers of sizes $n = \{512, 256, 128, 64, 32\}$. The residual network consists of 16 fully connected layers with an additional skip connection; each of size $n=32$. These are included to help preserve information as it passes through the network, and it stabilizes the training process, as discussed in more detail by Srivastava et al. \cite{srivastava2015highwaynetworks}. 
\\ \ \\
For each cell $I$ in the domain, the model is evaluated with an input feature tensor $\hat{\bm X}_I$ with dimensions $(2n_1+1) \times (2n_2+1) \times 9$ (for two dimensional cases). 
Here, $n_1$ and $n_2$ again denote the dimensions of the stencil. This tensor is constructed by evaluating the $9$ input feature channels $[\hat{u}_1, \hat{u}_2, \hat{u}^\text{lag}_{1}, \hat{u}^{\text{lag}}_2, \hat{S}_{11},\hat{S}_{12},\hat{S}_{22},q,w]$ at each stencil point surrounding cell $I$. To ensure a consistent input distribution during training, a \texttt{BatchNorm2D} \cite{ioffe2015batchnormalizationacceleratingdeep} layer is used, which normalizes the input channels to have zero mean and unit variance using running statistics.
The normalized features are then flattened to a vector and passed to the neural network, whose layers are defined as
\begin{eqnarray}
\text{EncLayer}_i(\hat{\bm X}) &=& \text{ReLU}({\bm W}_i\hat{\bm X} + {\bm b}_i),\\
\text{ResLayer}_j(\hat{\bm X}) &=& \text{ReLU}({\bm W}_j\hat{\bm X} + {\bm b}_j) + \hat{\bm X}\\
\textrm{and\ \ \ }
\text{OutLayer}(\hat{\bm X}) &=& {\bm W}_f\hat{\bm X} + {\bm b}_f,
\end{eqnarray}
where ${\bm W}_i$ and ${\bm b}_i$ are the weights and biases of each layer. At the end of the network, a linear layer is used to map the output to the dimensionless force term $\hat{\bm f}$. Here, the activation function is omitted to allow for an unrestricted output range. Finally, the output is re-dimensionalized by multiplication with the characteristic acceleration scale $\sqrt{k^*} \omega^*$ and transformed back to the global frame using the rotation operator ${\bm R}$ to obtain the force term $\bm f$.
\\ \ \\
To train the neural network, the  \texttt{PyTorch 2.0} library \cite{NEURIPS2019_9015,ansel2024pytorch} is employed. The \texttt{AdamW} optimizer is used to minimize the mean squared error loss function
\begin{eqnarray}
\label{e_mse_loss}
L = \frac{1}{N}\sum^{N}_\text{each cell $I$}|\hat{\bm f}(\hat{\bm X}_I^\textrm{ref}) - \hat{\bm f}^\textrm{ref}_I|^2,
\end{eqnarray}
where $N$ is the number of cells over all domains used in the training set. While $\hat{\bm f}(\hat{\bm X}_I^\textrm{ref})$ denotes the correction force evaluated by the neural network from the input $\hat{\bm X}^\textrm{ref}_I$ provided by the NLSS at grid cell $I$, $\hat{\bm f}^\textrm{ref}_I$ is the reference correction force vector at that location; obtained as described in section~\ref{s_intro}
and algorithm~\ref{alg:training_data}. 
\\ \ \\
To accelerate the training process, the loss function is evaluated in mini-batches of size 4096. The loss in prediction quality expected from such large batch sizes is not observed \cite{keskar_large-batch_2017}. All hyperparameters can be found in table~\ref{t_parameter_values}. During training, the neural network is monitored for over-fitting by evaluating the loss function $L$ on a separate validation set. At the end of the training process, the trained neural network with the lowest validation loss is used for integration into the RANS solver.


\subsection{Parameters, Algorithms and Code Availability}\label{sec:params}
The main objective of this section is to facilitate the reproduction of our results by specifying the necessary technical details. The values presented in table~\ref{t_parameter_values} sufficiently stabilized the solver without needlessly increasing the computational cost for obtaining steady state solutions ($\chi^\textrm{damp}$ and $\mu^\textrm{mem}$), resulted in smooth yet accurate evaluations of reference $\bm f$-values ($\chi_\textrm{max}$), resulted in appropriate shifted stencils ($c^\textrm{lag}$), led to large enough stencils with enough support ($n_1$, $n_2$ and $c_l$) and struck a balance between prediction speed and accuracy by only evaluating the NLSS model as often as needed ($T_{NLSS}$).
\\ \ \\
Algorithm~\ref{alg:training_data} describes the procedure used to obtain the data the NLSS model was trained on. Algorithm~\ref{alg:inference} details the prediction loop in which the trained NLSS model is used to correct a standard SIMPLE-based RANS solver. In addition, the full source code is available on our GitLab repository (\url{https://gitlab.ethz.ch/nlss/nlss-openfoam}).

\begin{table}[h]
	\centering
	\caption{Summary of parameters used in the study}
	\label{t_parameter_values}
	\begin{tabular}{l||l|l|l}
	\hline
	 & {\bf parameter} & {\bf value} & {\bf description} \\
	\hline
	\hline
	\multirow{4}{*}{\bf stencil} & $n_1, n_2$ & 7 & dimensions in $x$- and $y$-direction\\
	& $n_3$ & 0 & dimension in $z$-direction \\
	& $c_l$ & 1.5 & spatial support factor \\
	& $c_\text{lag}$ & 0.1 & Lagrangian shift factor \\

	\hline
	\multirow{5}{*}{k-$\omega$ \bf model} & $\beta^*$ & 0.09 &  \multirow{5}{*}{standard values - for a detailed description, refer to \cite{wilcox_turbulence_1998}}  \\
	& $\beta$ & 0.072 &  \\
	& $\gamma$ & 0.52 &  \\
	& $\alpha_k$ & 0.5 &  \\
	& $\alpha_\omega$ & 0.5 & \\
	\hline
	\multirow{6}{*}{\bf training} & $b$ & 4096 & batch size \\
	& $\chi_\text{max}$ & 5s$^{-1}$ & drift term rate \\
	& $N_\text{data}$ & $54$·10$^{3}$ & number of stencil-force pairs in training set \\
	& $N_\text{params}$ & $1.23$·10$^{6}$ & total number of network parameters \\
	& $\gamma$ & 3·10$^{-4}$ & learning rate \\
	& $\alpha$ & 1·10$^{-3}$ & weight decay \\
	\hline
	\multirow{3}{*}{\bf prediction} 	& $T_\text{NLSS}$ & 10 & evaluation interval \\
	& $\chi_\text{damp}$ & 0.5s$^{-1}$ & damping rate \\
	& $\mu_\text{mem}$ & 0.95 & memory factor \\
	\hline
	\end{tabular}
\end{table}

\begin{algorithm}[h]

\caption{Generation of training data}\label{alg:training_data}
\begin{algorithmic}[1]
	
\vspace {0.5em}
\State \textbf{Inputs:} Reference velocity field $\bar{\bm u}^{\text{high fidelity}}$ and corresponding RANS mesh $\{\bm{x}_i\}_{i=1}^N$
\State \textbf{Output:} Training data $\left\{(\bm{\hat X}_i^\text{ref}, \bm{\hat f}_i^\text{ref})\right\}_{i=1}^{2N}$
\vspace{0.5em}
\State $\bar{\bm u}_i^{\text{high fidelity}} \gets \bar{\bm u}^{\text{high fidelity}}(\bm{x}_i)~\text{\bf for}~i=1~\text{to}~N$ \Comment{linear interpolation to each RANS cell (see section~4.2)}
\vspace{0.5em}
\State $\bar{\bm u}^\text{ref}, k^\text{ref}, \omega^\text{ref}$ $\gets$ solution of RANS Eq. (14) with drift term $\chi(\bar{\bm u}_i^\text{high fidelity}-\bar{\bm u}_i)$
\State $\bm{f}^\text{ref} \gets \chi\left(\bar{\bm u}^{\text{high fidelity}} - \bar{\bm u}^\text{ref}\right)$
\vspace{0.5em}
\State $\phi_f \gets$ solution of $\nabla^2 \phi_f = \nabla \cdot \bm{f}^\text{ref}$ \Comment{Helmholtz projection}
\State $\bm{f}^\text{ref} \gets \bm{f}^\text{ref} - \nabla \phi_f$
\vspace{0.5em}
\For {$i=1$ to $N$} \Comment {for each RANS cell:}
	\State $\bar{\bm u}^\text{*}, k^\text{*}, \omega^\text{*}, \bm{x}^* \gets \bar {\bm u}^\text{ref}_i, k^\text{ref}_i, \omega^\text{ref}_i$, $\bm{x}_i$ \Comment {source reference quantities from $\bm{x}_i$}
	\vspace{0.5em}
	\State $\bm{x}_{I,J,K} \gets \text{StencilPoints}(\bar{\bm u}^\text{*}, k^\text{*}, \omega^\text{*}, \bm{x}^*)$ \Comment{compute stencil points (see Eq.~(8))}
	\State $\bm{X} \gets \text{Sample}(\bm{x}_{I,J,K};\bar {\bm u}^\text{ref}, k^\text{ref}, \omega^\text{ref})$ \Comment{sample required fields at stencil points}
	\vspace{0.5em}
	\State $\bm{\hat X}_i^\text{ref} \gets \text{Transform}(\bm{X}; \bar{\bm u}^\text{*}, k^\text{*}, \omega^\text{*}, \bm{x}^*)$ \Comment{transform and normalize features (see Eqs.~(9-12))}
	\State $\bm{\hat f}_i^\text{ref} \gets \text{Transform}(\bm{f}^\text{ref}_i; \bar{\bm u}^\text{*}, k^\text{*}, \omega^\text{*}, \bm{x}^*)$ \Comment{transform and normalize  target force (see Eq.~(13))}
	\vspace{0.5em}
	\State $\bm{\hat X}_{N+i}^\text{ref} \gets \text{Mirror}(\bm{\hat X}_i^\text{ref})$ \Comment{data augmentation (see Eqs.~(32-37))}
	\State $\bm{\hat f}_{N+i}^\text{ref} \gets \text{Mirror}(\bm{\hat f}_i^\text{ref})$
\EndFor\\
\vspace{0.5em}
\Return $\left\{(\bm{\hat X}_i^\text{ref}, \bm{\hat f}_i^\text{ref})\right\}_{i=1}^{2N}$ \Comment{return dataset containing both the original and mirrored stencil-force pairs}
\end{algorithmic}
\end{algorithm}
\begin{algorithm}[h]
\caption{Prediction loop}\label{alg:inference}
\begin{algorithmic}[1]
\vspace {0.5em}
\State \textbf{Inputs:} RANS mesh $\{\bm{x}_i\}_{i=1}^N$, uncorrected RANS solution, trained model $\mathcal{M}: \hat{\bm X} \mapsto \hat{\bm f}$
\State \textbf{Output:} NLSS corrected RANS solution $\bar{\bm u}^\text{NLSS}$
\vspace{0.5em}
\State $\bar{\bm u}^1 \gets \bar{\bm u}^\text{RANS}$ \Comment{initial guess from uncorrected RANS solution}
\State $k^1 \gets k^\text{RANS}$
\State $\omega^1 \gets \omega^\text{RANS}$ 
\State $\bm{\bar u}^{\text{MA},1} \gets \bar{\bm u}^1$ \Comment{moving average initialization}
\State $n \gets 1$
\vspace{0.5em}
\While {not converged}
\If {$n$ is divisible by $T_\text{NLSS}$} \Comment {only evaluate the correction force every $T_\text{NLSS}$-th iteration}
\For {$i=1$ to $N$} \Comment{for each RANS cell:}
	\State $\bar{\bm u}^*, k^*, \omega^*, \bm{x}^* \gets \bar{\bm u}^n_i, k^n_i, \omega^n_i, \bm{x}_i$ \Comment{source reference quantities from $\bm{x}_i$}
	\vspace{0.5em}
	\State $\bm{x}_{I,J,K} \gets \text{StencilPoints}(\bar{\bm u}^*, k^*, \omega^*, \bm{x}^*)$ \Comment{compute stencil points (see Eq.~(8))}
	\State $\bm{X} \gets \text{Sample}(\bm{x}_{I,J,K};\bar{\bm u}^n, k^n, \omega^n)$ \Comment{sample required fields at stencil points}
	\State $\bm{\hat X} \gets \text{Transform}(\bm{X}; \bar{\bm u}^*, k^*, \omega^*, \bm{x}^*)$ \Comment{transform and normalize features (see Eqs.~(9-12))}
	\vspace{0.5em}
	\State $\bm{\hat f} \gets \mathcal{M}(\bm{\hat X}_i)$ \Comment{evaluate the model}
	\State $\bm{f}_i \gets \text{Transform}^{-1}(\bm{\hat f}_i; \bar{\bm u}^*, k^*, \omega^*, \bm{x}^*)$ \Comment{inverse transform and redimensionalize force (see Eq.~(13))}
\EndFor
\vspace{0.5em}
\State $\phi_f \gets$ solution of $\nabla^2 \phi_f = \nabla \cdot \bm{f}^\text{ref}$ \Comment{Helmholtz projection}
\State $\bm{f} \gets \bm{f} - \nabla \phi_f$

\Else
\State reuse $\bm{f}$ from previous iteration
\EndIf
\vspace{0.5em}
\State $\bm{s}_u \gets \bm{f} + \chi^\text{damp}(\bar{\bm u}^{\text{MA},n} - \bar{\bm u}^n)$ \Comment{momentum source term (correction and damping)}

\State $\bar{\bm u}^{n+1}, k^{n+1}, \omega^{n+1} \gets \text{SIMPLEStep}(\bar{\bm u}^n, k^n, \omega^n; \bm{s}_u)$ \Comment{perform a single SIMPLE step}

\State $\bm{\bar u}^{\text{MA},n+1} \gets (1-\mu_\text{mem}) \bar{\bm u}^{n+1} + \mu_\text{mem} \bar{\bm u}^{\text{MA},n}$ \Comment{update moving average}
\vspace{0.5em}
\State $n \gets n+1$
\EndWhile\\
\vspace{0.5em}
\Return $\bar{\bm u}^n$ \Comment{return the converged NLSS-corrected RANS solution}
\end{algorithmic}
\end{algorithm}

\FloatBarrier

\section*{Acknowledgement}
The authors express their thankfulness to Daniel W. Meyer for constructive discussions and proofreading, to Justin Plogmann and Oliver Brenner for their help with using {\em OpenFOAM}, as well as for constructive discussions, and to Hossein Gorji for discussions regarding the current state of research in the field. The authors also thank the reviewers; their constructive comments helped to significantly improve the clarity of the paper. Further, while having written the paper independently and in own words, the authors confirm that they employed generative artificial intelligence technologies (i.e., {\em ChatGPT~4o}) for linguistic improvement of the text.

\bibliographystyle{unsrt}
\bibliography{paper_nlss_submit.bib}

\end{document}